# Effects of Switching Behavior for the Attraction on Pedestrian Dynamics

**Jaeyoung Kwak[1]*, Hang-Hyun Jo[2,3], Tapio Luttinen[1], Iisakki Kosonen[1]**

**1** Department of Civil and Environmental Engineering, Aalto University, Espoo, Finland, **2** BK21plus Physics Devision and Department of Physics, Pohang University of Science and Technology, Pohang, Republic of Korea, **3** Department of Computer Science, Aalto University, Espoo, Finland

* jaeyoung.kwak@aalto.fi

**Citation:** Kwak J, Jo H-H, Luttinen T, Kosonen I (2015) Effects of Switching Behavior for the Attraction on Pedestrian Dynamics. PLoS ONE 10(7): e0133668. doi:10.1371/journal.pone.0133668

**Editor:** Frederic Amblard, University Toulouse 1 Capitole, FRANCE

**Received:** October 9, 2014

**Accepted:** June 30, 2015

**Published:** July 28, 2015

**Data Availability Statement:** All relevant data are within the paper.

**Funding:** JK was supported by Aalto University School of Engineering Doctoral Program. HJ was supported by Aalto University postdoctoral program and by Mid-career Researcher Program through the National Research Foundation of Korea (NRF) grant funded by the Ministry of Science, ICT and Future Planning (2014030018), and Basic Science Research Program through the National Research Foundation of Korea (NRF) grant funded by the Ministry of Science, ICT and Future Planning (2014046922). The funders had no role in study design, data collection

## Abstract

Walking is a fundamental activity of our daily life not only for moving to other places but also for interacting with surrounding environment. While walking on the streets, pedestrians can be aware of attractions like shopping windows. They can be influenced by the attractions and some of them might shift their attention towards the attractions, namely switching behavior. As a first step to incorporate the switching behavior, this study investigates collective effects of switching behavior for an attraction by developing a behavioral model. Numerical simulations exhibit different patterns of pedestrian behavior depending on the strength of the social influence and the average length of stay. When the social influence is strong along with a long length of stay, a saturated phase can be defined at which all the pedestrians have visited the attraction. If the social influence is not strong enough, an unsaturated phase appears where one can observe that some pedestrians head for the attraction while others walk in their desired direction. These collective patterns of pedestrian behavior are summarized in a phase diagram by comparing the number of pedestrians who visited the attraction to the number of passersby near the attraction. Measuring the marginal benefits with respect to the strength of the social influence and the average length of stay enables us to identify under what conditions enhancing these variables would be more effective. The findings from this study can be understood in the context of the pedestrian facility management, for instance, for retail stores.

## Introduction

Collective dynamics of interactions among individuals is one of the current central topics in the field of interdisciplinary physics. In an attempt to quantify self-organized phenomena in the real world, various topics have been studied such as opinion formation [1, 2], spread of disease [3], and pedestrian dynamics [4]. The study of pedestrian dynamics has generated considerable research interests in the physics community and various models have been proposed in order to describe fascinating collective phenomena, such as cellular automata (CA) approaches [5, 6], force-based models [7, 8], and heuristic-based models [9, 10]. The CA approaches discretize walking space into two dimensional lattices where each cell can have at most one agent.

and analysis, decision to publish, or preparation of the manuscript.

**Competing Interests:** The authors have declared that no competing interests exist.

Agents move between cells according to the probability of movement. The CA approaches offer simple and efficient computation, but yield limited accuracy on describing pedestrian movements due to the nature of discretized time and space. Force-based models present internal motivation and external stimuli as force terms and describe the movement of pedestrians by summing these force terms in continuous space. Although force-based models can describe pedestrian behaviors in detail, they are computationally expensive. Heuristic-based models predict pedestrian movements based on a set of rules defined for different situations. The heuristics-based models can intuitively simulate pedestrian movements for specific situations, but at the same time they might be able to mimic only a limited number of situations. Among these models, force-based models have been remarkably extended by improving behavioral aspects of pedestrian movements including repulsive interactions [11–13], collision prediction [14], pedestrian group behavior [15–17], and attractive interactions between pedestrians and attractions [18].

Numerical studies in Ref. [18] examined attractive interactions between pedestrians and attractions by appending the attractive force toward the attractions. Although their extended social force model exhibited various collective patterns of pedestrian movements, their study did not explicitly take into account selective attention. The selective attention is a widely recognized behavioral mechanism by which people can focus on tempting stimuli and disregard uninteresting ones [19, 20]. While walking on the streets, pedestrians can be aware of attractions like shopping windows and street performances. When such attractions come into sight, individuals can make decisions between moving in their initially planned directions and stopping by the attractions. This behavior can be called switching behavior, meaning that the individuals can shift their attention towards the attractions [18].

In reality, such switching behavior is likely to be influenced by the behavior of others. For instance, Milgram *et al.* [21] and Gallup *et al.* [22] reported that passersby on the streets were more likely to pay attention to stimulus crowds as the size of the crowd was increased. They noted that the size of the crowd was associated with the individual time spent looking at the stimulus crowd and social context. In the marketing area, it is widely believed that longer store visit duration and stronger social interactions are likely to attract more shoppers to stores. Having more visitors in the stores can affect one's information processing by raising awareness of merchandise displays in stores, and consequently the individual becomes prone to make more purchases [23, 24]. Therefore, marketing strategies have focused on increasing the length of store visit duration and the strength of social interactions [25].

In order to incorporate the switching behavior, this study constructs a simple probability model in which the preference for the attraction depends on the number of people who have already joined the attraction. By means of numerical simulations, the effects of the social influence and the average length of stay at the attraction are examined, and illustrated with a phase diagram.

The remainder of this paper is organized as follows. We first describe the proposed switching behavior model in the next section. Then we present its numerical simulation results with a phase diagram in the results section. Finally, we discuss the findings of this study in the section following the results.

# Methods

## Switching behavior

By the analogy with sigmoidal choice rule [21, 22, 26], the probability that a pedestrian joins an attraction point $P_a$ can be formulated based on the number of pedestrians who have already joined $N_a$ and the number of pedestrians not stopping by the attraction $N_0$:

$$P_a = \frac{sN_a}{N_0 + sN_a}. \quad (1)$$

This indicates that the joining probability increases as $N_a$ grows. Here $s > 0$ is the strength of the social influence that can be also understood as pedestrians' awareness of the attraction. When $s$ is small, individuals pay little attention to others' choice. For large $s$, the joining probability is more likely to be influenced by the number of people who have already joined the attraction rather than the number of people not joining the attraction. However, note that the joining probability cannot be determined when there is nobody within the range of perception, i.e., $N_a = N_0 = 0$. In order to avoid such indeterminate case, we introduce positive constants $K_a$ for joining the attraction and $K_0$ for not stopping by the attraction as baseline values of $N_a$ and $N_0$, respectively, as follows:

$$P_a = \frac{s(N_a + K_a)}{(N_0 + K_0) + s(N_a + K_a)}. \quad (2)$$

According to previous studies [21, 22, 25], here we postulated that the strength of social influence can be different for different situations and can be controlled in the presented model. After joining the attraction, the individual will then stay near the attraction for an exponentially distributed time with an average of $t_d$, similar to previous works [7, 22, 27].

## Pedestrian movement

According to the social force model [7], the velocity $\vec{v}_i(t)$ of pedestrian $i$ at time $t$ is given by the following equation:

$$\frac{\mathrm{d}\vec{v}_i(t)}{\mathrm{d}t} = \frac{v_d\vec{e}_i - \vec{v}_i(t)}{\tau} + \sum_{j\neq i}\vec{f}_{ij} + \sum_{B}\vec{f}_{iB}. \quad (3)$$

Here the first term on the right-hand side indicates the driving force describing the tendency of pedestrian $i$ moving toward his destination with the desired speed $v_d$ and an unit vector $\vec{e}_i$ pointing to the desired direction. The relaxation time $\tau$ controls how fast pedestrian $i$ adapts its velocity to the desired velocity. The repulsive force terms $\vec{f}_{ij}$ and $\vec{f}_{iB}$ reflect his tendency to keep certain distance from other pedestrian $j$ and the boundary $B$, e.g., wall and obstacles. Based on previous studies [7, 12, 18], the repulsive force between pedestrians $i$ and $j$ is specified by summing the gradient of repulsive potential with respect to $\vec{d}_{ij} \equiv \vec{x}_j - \vec{x}_i$, and the friction force, $\vec{g}_{ij}$:

$$\vec{f}_{ij} = -\nabla_{\vec{d}_{ij}}\left[C_p l_p \exp\left(-\frac{b_{ij}}{l_p}\right)\right] + \vec{g}_{ij}. \quad (4)$$

$C_p$ and $l_p$ are the strength and the range of repulsive interaction between pedestrians $i$ and $j$. Here $b_{ij} = \frac{1}{2}\sqrt{(\|\vec{d}_{ij}\| + \|\vec{d}_{ij} - \vec{y}_{ij}\|)^2 - \|\vec{y}_{ij}\|^2}$ is the effective distance between pedestrians $i$ and $j$ by assuming their relative displacement $\vec{y}_{ij} \equiv (\vec{v}_j - \vec{v}_i)\Delta t$ with the stride time $\Delta t$ [12]. The interpersonal friction $\vec{g}_{ij} = kh(r_{ij} - d_{ij})\vec{e}_{ij}$ becomes effective when the distance $d_{ij} = \|\vec{d}_{ij}\|$ is smaller than the sum $r_{ij} = r_i + r_j$ of their radii $r_i$ and $r_j$, where $k$ is the normal elastic constant and $\vec{e}_{ij}$ is an unit vector pointing from pedestrian $j$ to $i$. The function $h(x)$ yields $x$ if $x > 0$, while it gives 0 if $x \leq 0$. The repulsive force from boundaries is denoted as $\vec{f}_{iB} = C_b\exp(-d_{iB}/l_b)\vec{e}_{iB}$, where $d_{iB}$ is the perpendicular distance between pedestrian $i$ and wall, and $\vec{e}_{iB}$ is the unit vector pointing from the wall $B$ to the pedestrian $i$. $C_b$ and $l_b$ denote the strength and the range of repulsive interaction from boundaries.

## Numerical simulation setup

Each pedestrian is modeled by a circle with radius $r_i = 0.25$ m. $N = 100$ pedestrians move in a corridor of length 30 m and width 6 m with periodic boundary condition in the horizontal direction. They move with desired speed $v_d = 1.2$ m/s and with relaxation time $\tau = 0.5$ s, and their speed cannot exceed $v_{\max} = 2.0$ m/s. The desired direction points from the left to the right boundary of the corridor for one half of population and the opposite direction for the other half. The parameters of the repulsive force terms are given based on previous works: $C_p = 3$, $l_p = 0.2$, $k = 62.5$, $C_b = 10$, and $l_b = 0.2$ [7, 10, 12, 18].

The joining probability (Eq (2)) is updated with the social force model (Eq (3)) for each simulation time step of 0.05 s. The individual can decide whether he will join the attraction when the attraction comes into his perception range $R_i = 10$ m. Once the individual decides to join the attraction, then he shifts his desired direction vector $\vec{e}_i$ toward the attraction. The attraction is placed at the center of lower wall, i.e., at the distance of 15 m from the left boundary of the corridor. For the sake of simplicity, $K_a$ and $K_0$ are set to be 1, meaning that both options are equally attractive when the individual would see nobody within his perception range. An individual is counted as an attending pedestrian if his efficiency of motion $E_i = (\vec{v}_i \cdot \vec{e}_i)/v_d$ is lower than 0.05 within a range of 3 m from the center of the attraction after he decided to join there. Here the efficiency of motion indicates how much the driving force contributes to the pedestrian motion with a range from 0 to 1 [18, 28]. $E_i = 1$ implies that the individual is walking towards his destination with the desired velocity while lower $E_i$ indicates that an individual is distracted from his initial destination because of the attraction.

# Results and Discussion

## Feasibility zone

The simulation results show different patterns of pedestrian movements depending on the strength of the social influence $s$ and the average length of stay $t_d$. If the social influence is weak, one can define an unsaturated phase where some pedestrians move towards the attraction while others walk in their desired directions (see Fig 1(a)). For large values of $s$, a saturated phase can be defined, in which every pedestrian near the attraction heads for the attraction (see Fig 1(b)).

We focus on the behavior of the number of pedestrians who visited the attraction $N_v$, out of pedestrians within a range of $R_a = 10$ m from the center of the attraction, denoted by $N_p$. Then the number of pedestrians not visiting the attraction is $N_p - N_v$. Quantifying the behavior of $N_v$ is useful because it reflects the level of crowding and helps facility managers to estimate the cost of making the accommodation necessary for visitors. Fig 2 shows how $N_v$ depends on the strength of social influence $s$ and the average length of stay $t_d$. For a given $t_d$, $N_v$ increases according to $s$, indicating that more pedestrians are distracted from their initial desired velocity due to others' choice on the attraction, see the left panel of Fig 2. Furthermore, $N_v$ curves rapidly increase as $t_d$ grows, meaning that the larger $t_d$, the smaller $s$ is needed to attract the majority of pedestrians (see the right panel of Fig 2). The increasing behavior of $N_v$ in the right panel of Fig 2 shows a similar pattern to that in the left panel.

Although $N_v$ enables us to evaluate the attraction influence on pedestrians, quantifying marginal benefits of facility improvements can indicate the effective ways of enhancing the attraction influence on pedestrians. The increase of $N_v$ can be understood as the benefit of the facility improvements in the sense that the attraction is likely to have more potential customers with higher $N_v$. Firstly, $\Delta N_v/\Delta s$ defines the marginal benefit with respect to the change of *s*, which can be calculated as the first derivative of $N_v$ curves in the left panel of Fig 2. In a similar way, we can also evaluate $\Delta N_v/\Delta t_d$ from the first derivative of $N_v$ curve in the right panel of Fig 2, which defines the marginal benefit with respect to the average length of stay $t_d$. As indicated in Fig 3(a) and 3(b), each marginal benefit curve increases up to a certain level and then decreases. For higher $t_d$ and *s*, the values of the marginal benefits are sensitive to the variations of *s* and $t_d$ for low values of *s* and $t_d$, respectively. Thus, one can find certain ranges or windows of *s* and $t_d$, where the effective improvements of $N_v$ can be achieved. The variations of *s* and $t_d$ are considered as effective if $\Delta N_v/\Delta s$ is larger than 30 and if $\Delta N_v/\Delta t_d$ is larger than 0.3. Feasible conditions of the effective improvements can be further illustrated in the parameter space of *s* and $t_d$ by means of $\Delta N_v/\Delta s$ and $\Delta N_v/\Delta t_d$, as shown in Fig 4. The parameter region can be called a feasibility zone of the effective improvements, indicating that higher increase of $N_v$ is expected with smaller increase of *s* and $t_d$. Other than the feasibility zone, the impact of changing those variables are insignificant.

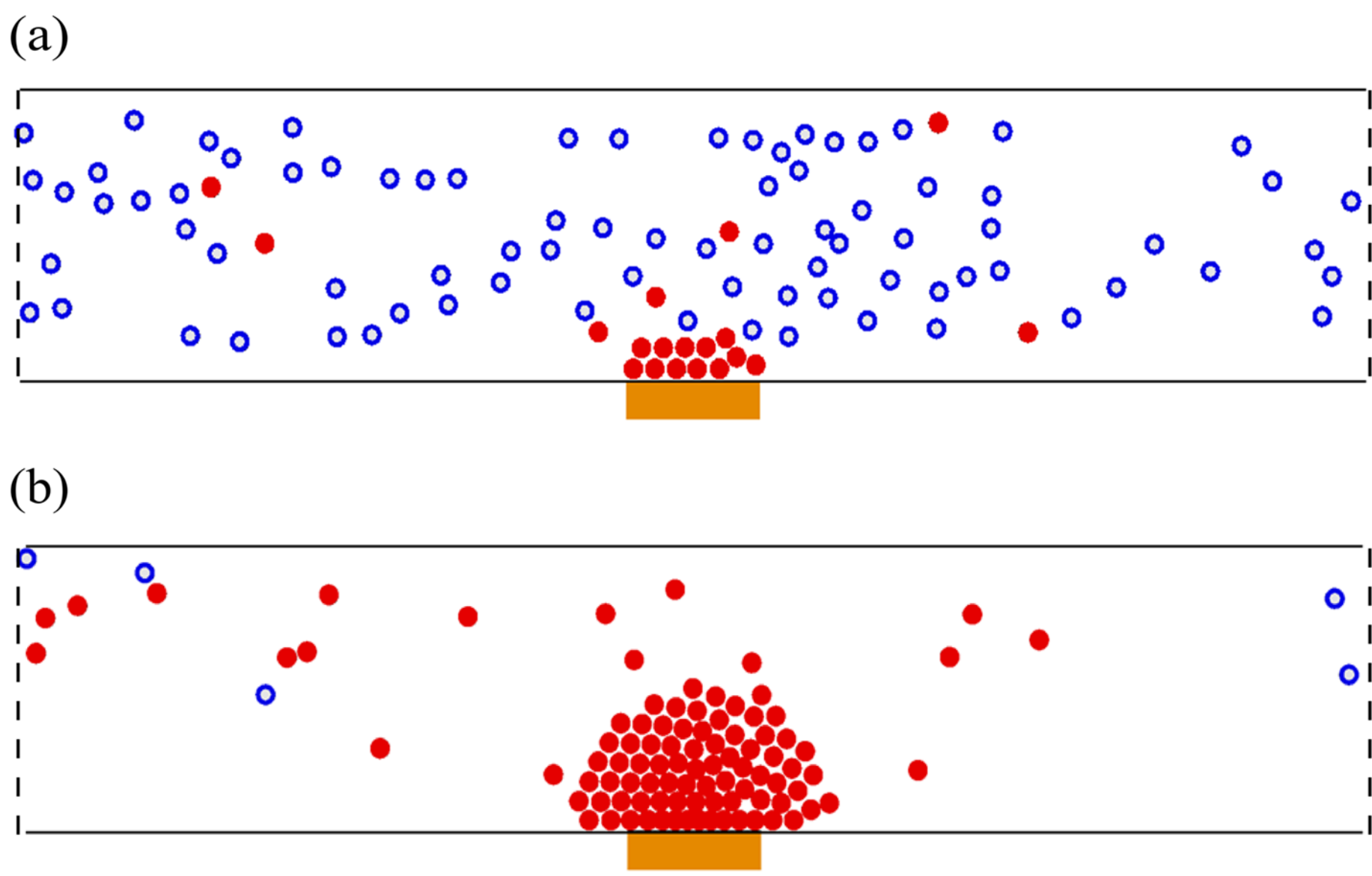

**Fig 1. Representative snapshots of numerical simulations with various values of the social influence *s* and the average length of stay $t_d$.** The attraction, depicted by an orange rectangle, is located at the center of the lower wall. Filled red and hollow blue circles depict the pedestrians who have and have not visited the attraction, respectively. Two phases are observed: (a) Unsaturated phase in the case of $s = 0.4$ and $t_d = 30$ s, in which some pedestrians have visited the attraction while others walk in their desired directions. (b) Saturated phase in the case of $s = 1.5$ and $t_d = 60$ s, where all the pedestrians around the attraction, within a range of $R_a = 10$ m, have visited the attraction.

## Phase diagram

The parameter space of *s* and $t_d$ is divided into two regions according to the value of $N_v$. For each value of $t_d > 43$ s, $N_v$ increases as *s* increases, and then finally reaches $N_p$ at a critical value of *s*, i.e., $s_c$. This indicates the transition from the unsaturated phase to the saturated phase (see Fig 4). For the region of $t_d < 43$ s, $N_v$ increases as *s* grows but does not reach to its maximum allowed value even for the large values of *s*, as can be seen from Fig 2(a) and 2(c). The average length of stay $t_d$ is not long enough, so it fails to obtain sufficient amount of visitors. One can also observe that $s_c$ substantially decreases for $t_d \leq 120$ s and appears to stay around 0.2 when $t_d$ is larger than 180 s. It is reasonable to suppose that the departure rate of attending pedestrians is positively associated with the reciprocal of $t_d$ while the arrival rate of joining pedestrians is linked to $s$. Accordingly, we can infer that the impact of the departure rate on $N_v$ is dominated by that of the arrival rate when $t_d$ is large, so the marginal impact of increasing $t_d$ becomes less notable for larger $t_d$.

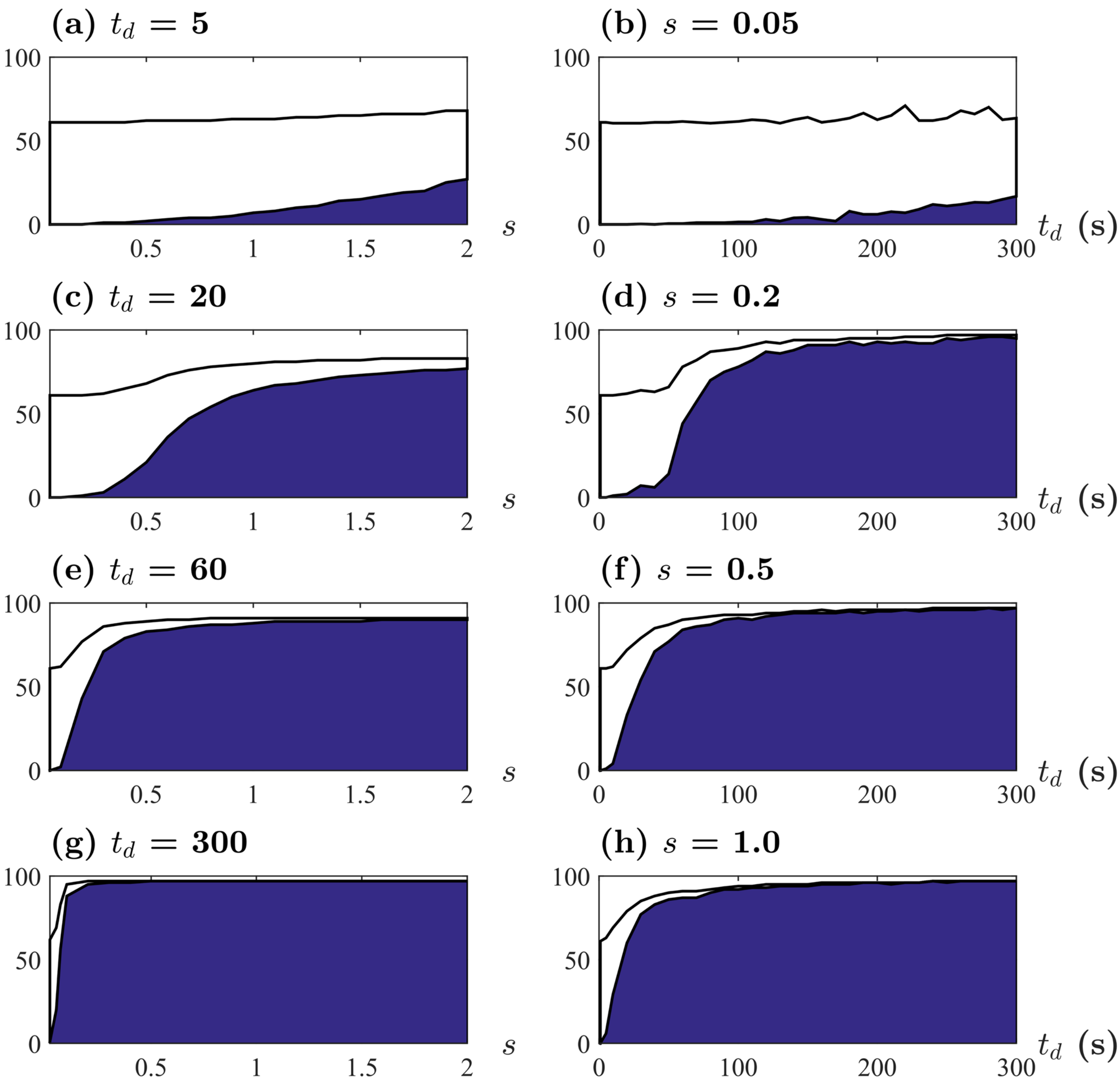

**Fig 2. Numerical results of $N_v$ and $N_p$.** Filled blue area indicates the number of pedestrians who visited the attraction $N_v$ and hollow area bounded by solid lines corresponds to the number of pedestrians not visiting the attraction $N_p - N_v$. Top subplots present the cases of small $s$ (a) and small $t_d$ (b) while bottom subplots for large $s$ (g) and large $t_d$ (h). It is observed that higher values of $t_d$ and $s$ appear to yield rapid growth of $N_v$ for smaller intervals as $s$ and $t_d$ increase, respectively.

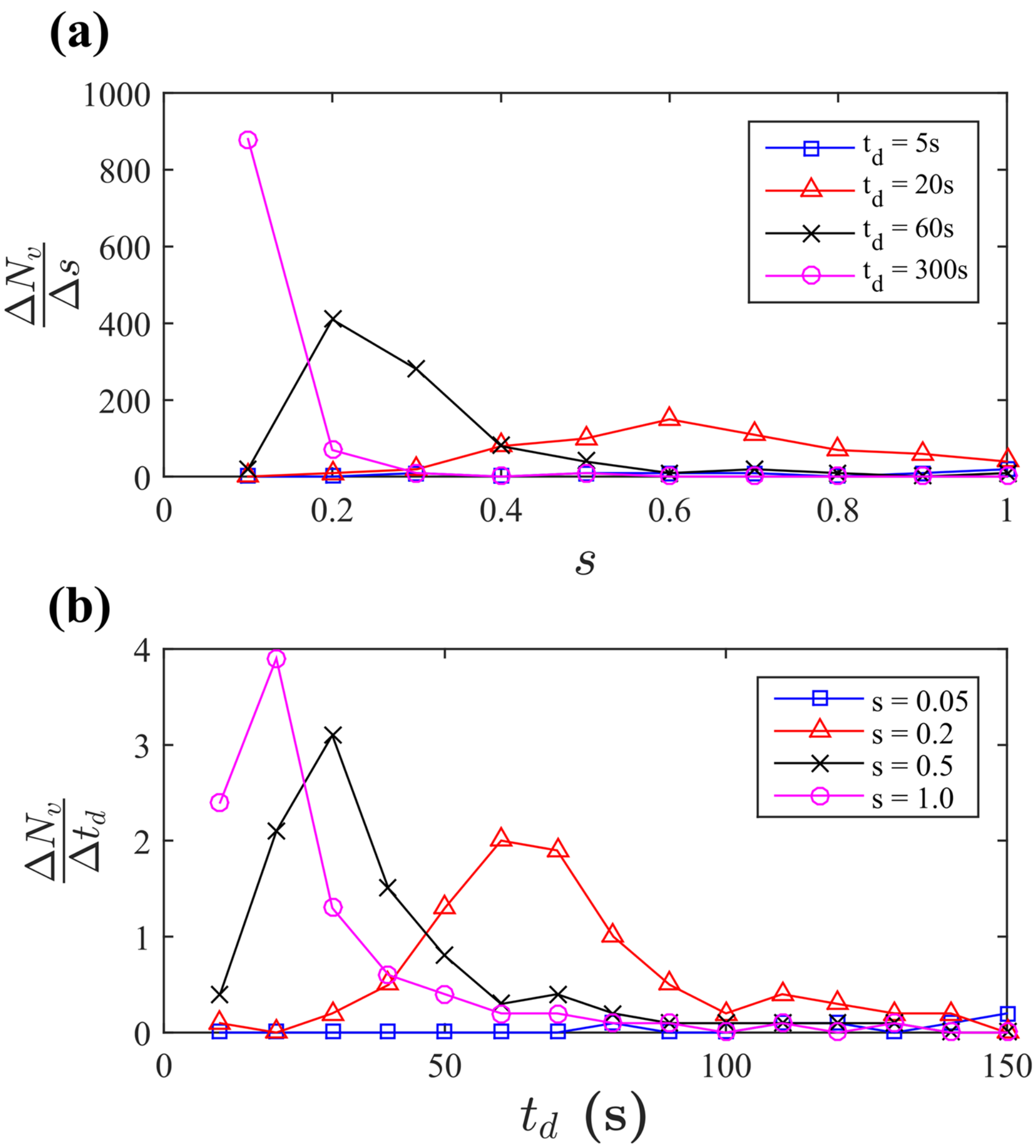

**Fig 3. Numerical results of the marginal benefits, Δ*N*$_v$/Δ*s* (a) and Δ*N*$_v$/Δ*t*$_d$ (b).** The marginal benefits increase up to a certain level and then decrease for given $t_d$ and $s$, respectively. Interestingly, the marginal benefits are sensitive to the variations of $s$ and $t_d$ within the bounded intervals, showing that effective improvements of these variables can be achieved. Different symbols represent the different values of $t_d$ and $s$.

## Conclusion

In order to examine the collective effects of the switching behavior, this study has developed a behavioral model of pedestrians' joining an attraction. A phase diagram with different collective patterns of pedestrian behavior is presented. The phases are identified by comparing the number of pedestrians who visited the attraction $N_v$ to the number of passersby near the attraction $N_p$. For strong social influence, the saturated phase appears where all pedestrians were enticed by the attraction, so all of them have visited the attraction. When the social influence is weak, the unsaturated phase is observed where the attraction is not captivating enough to entice all the pedestrians. Based on the numerical simulation results, feasible conditions of the effective improvements have been identified in terms of the marginal benefits. It is noted that the results are qualitatively similar to the cases of other $K_a$ and $K_0$ values. When $K_0$ is set to larger than $K_a$, the individual will have a higher probability of walking towards the initial destination than the case of joining the attraction. On the other hand, if $K_a$ is larger than $K_0$, the choice probability is increased. This study therefore presents a new approach to quantifying the collective effects of switching individuals' attention to the attraction depending on the strength of the social influence $s$ and the average length of stay $t_d$.

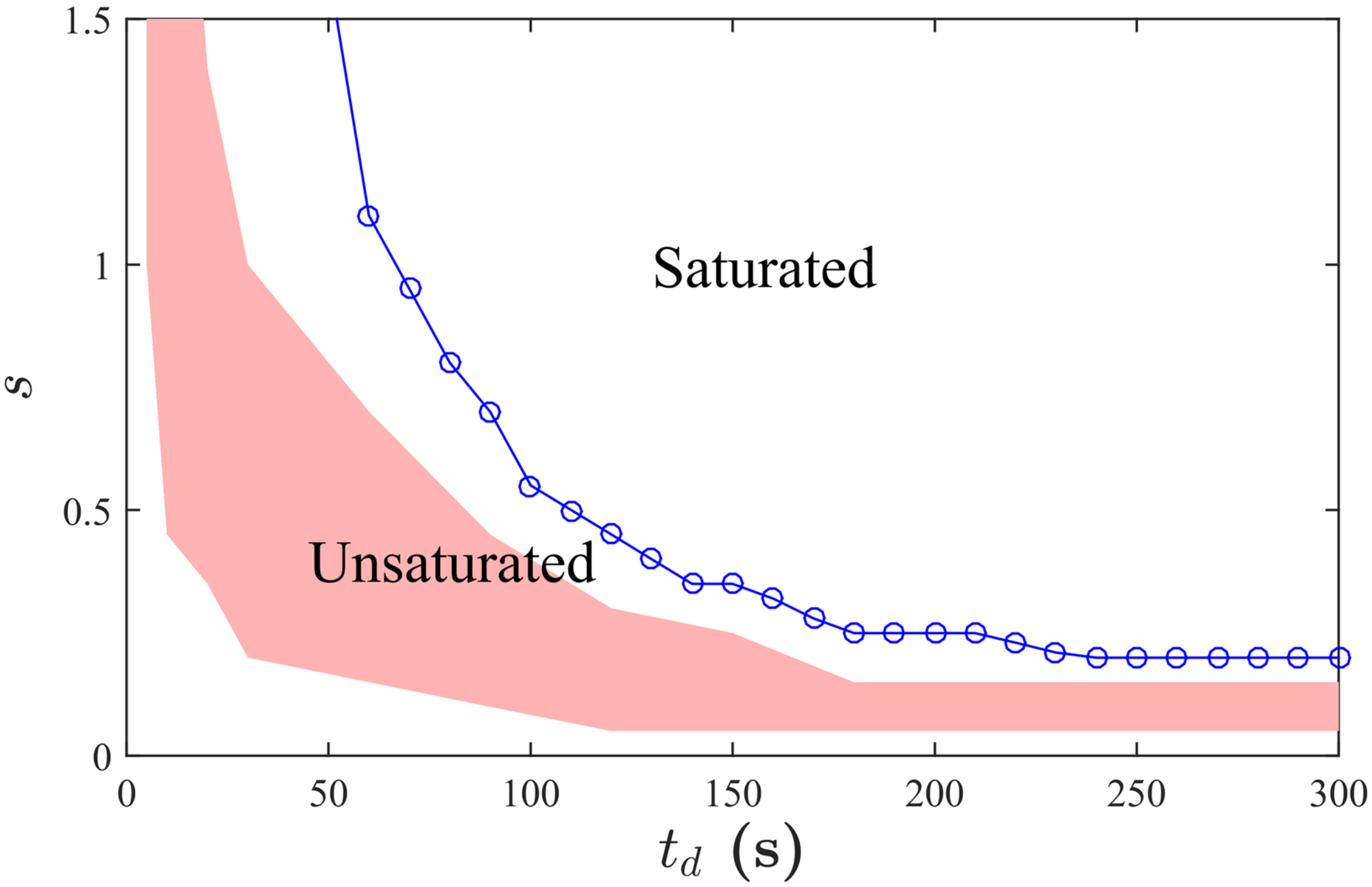

**Fig 4. A feasibility zone of effective improvements and an envelope of the saturated phase.** The red shade area depicts a feasibility zone of the effective improvements characterized in terms of the marginal benefits $\Delta N_v/\Delta s$ and $\Delta N_v/\Delta t_d$. The blue solid line with circle symbols represents the envelope of the saturated phase at which $N_v = N_p$. The saturated phase indicates that all the pedestrians have visited the attraction, while the unsaturated phase represents that not all pedestrians are enticed by the attraction.

The findings from this study can be contextualized in the domain of pedestrian facility management. The number of pedestrians who visited the attraction $N_v$ is relevant to store traffic which makes it possible to assess the attractiveness of the facility and predict the number of potential customers [29, 30]. The appearance of the saturated phase can be interpreted as a thorough dissemination of important information to the visitors, indicating that all the visitors are informed of events, for instance conferences and campaigns. The social influence $s$ can be understood as pedestrians' awareness of the attraction. For instance, an attraction in a busy train station tends have a low $s$ value so a few number of pedestrians might gather around the attraction. If there is an artwork produced by a famous artist in a museum (i.e., moderate $s$) and the value of $t_d$ is large, the artwork is likely to attract a large group of passersby around the artwork. On the other hand, when $t_d$ value of the artwork is small, it might fail to attract many people. In practice, for pedestrian facilities such as stores and museums, there are likely to exist costs associated with increasing the strength of social influence $s$ and the average length of stay $t_d$. Facility managers can think of increasing $t_d$ by making the pedestrian attractions more comfortable, so they might expect that visitors would stay longer. Although the strength of the social influence seems to be given for specific situations such as different time of day or locations, the managers can attempt to increase $s$ by changing the layouts of the facilities or changing pedestrian traffic density near the attractions. With the feasibility zone of effective investments, one can identify under what conditions facility improvements would be more effective.

As a first step to study the collective effects of switching behavior, a simple scenario has been considered. Based on the findings from this study, the following recommendations can be made for future research. First, a future study is recommended for the case with multiple attractions in order to consider more realistic situations such as a shopping street having several stores. An important question for the future study is to understand the spatio-temporal dynamics of pedestrians near the attractions. The future study will specifically examine the effects of increasing the number of attractions and the size of pedestrian clusters at the attractions. Second, one can take into account heterogeneous properties of attractions and pedestrians such as the strength of social influence and the average length of stay. In addition, different pedestrians reveal diverse characteristics in terms of desired speed, destinations, and preference on attractions. While this paper only considers homogeneous properties of pedestrians and attractions for the purpose of understandability, considering those heterogeneous properties are needed for immediate applications of this study. Finally, the presented model can be extended by incorporating various aspects of pedestrian behavior. For instance, we can postulate that pedestrians have time budget, so they evaluate the attractiveness and the cost of time due to joining the attractions.

## Acknowledgments

Jaeyoung Kwak would like to thank Aalto University School of Engineering Doctoral Program for financial support. He is grateful to CSC–IT Center for Science, Finland for providing computational resources. Hang-Hyun Jo gratefully acknowledges financial support by Aalto University postdoctoral program and by Mid-career Researcher Program through the National Research Foundation of Korea (NRF) grant funded by the Ministry of Science, ICT and Future Planning (2014030018), and Basic Science Research Program through the National Research Foundation of Korea (NRF) grant funded by the Ministry of Science, ICT and Future Planning (2014046922).

## Author Contributions

Conceived and designed the experiments: JK HJ. Performed the experiments: JK. Analyzed the data: JK HJ. Wrote the paper: JK HJ TL IK.